\documentclass[a4paper,11p$_T$]{article}
\usepackage{pos}

\title{Jet Energy Scale and Resolution Measurements in CMS}

\author*[a]{Garvita Agarwal}

\affiliation[a]{University at Buffalo - State University of New York,\\
  210 Talbert Hall, Buffalo, NY 14260, United States}

\emailAdd{garvitaa@buffalo.edu}

\abstract{Measurements of jet energy scale (JES) and resolution (JER) are presented, based on the legacy reconstruction of 13 TeV proton-proton collision data collected by the CMS experiment during the LHC Run 2 period from 2016-2018. Precision measurement of JES is of the utmost importance for the vast majority of physics measurements and searches at CMS. The high pileup, a harsh radiation environment, and time-dependent variations in detector response and calibration, all make precision JES measurement a challenging task. We present in-situ derivations of JES and JER based on CMS Run 2 data, as well as on simulated samples using various advanced techniques.}

\FullConference{%
  41st International Conference on High Energy physics - ICHEP2022\\
  6-13 July, 2022\\
  Bologna, Italy
}

\begin{document}
\maketitle

\section{Introduction}

Quarks and gluons are produced abundantly in high-energy proton-proton collisions at the LHC.
Color confinement causes the quarks and gluons to fragment and hadronize into a spray of stable particles ($c\tau > 1~\mathrm{cm}$) called jets.
Proper calibration of jets, i.e. ensuring that the energy and momentum of the reconstructed jet matches that of the quark/gluon-initiated jet, is extremely crucial for Standard Model (SM) measurements and Beyond Standard Model (BSM) searches.
Furthermore, the achieved calibration precision defines the accuracy of many measurements and the sensitivity of searches in CMS \cite{Chatrchyan:1129810} such as in the very precise measurement of the top quark mass \cite{CMS-PAS-TOP-20-008}.


\begin{figure}[!ht]
    \centering
    \includegraphics[scale=0.45]{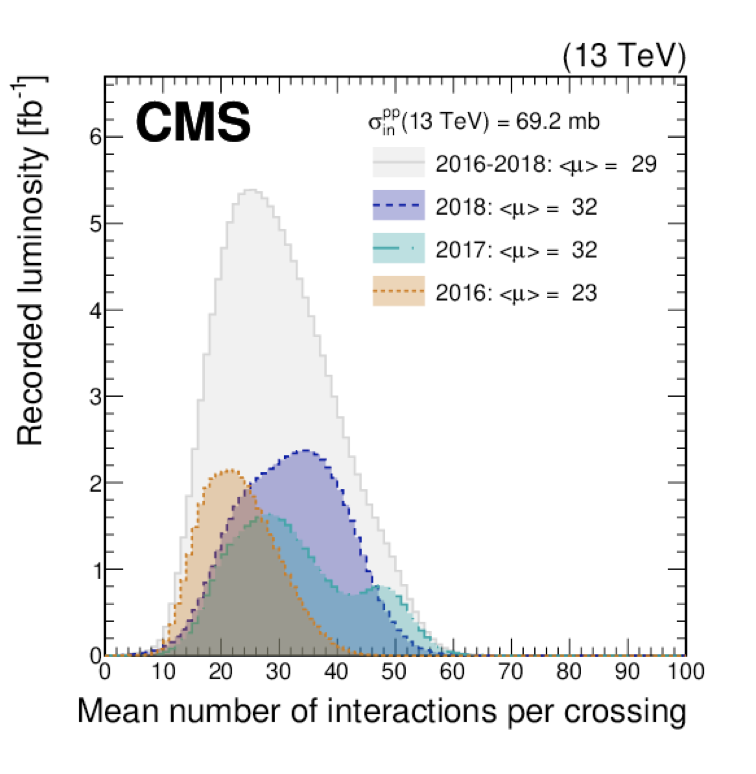}
    \caption{Pileup distribution in data for proton-proton collisions observed during Run 2. \cite{CMS-PAS-JME-18-001}}
    \label{fig:run2_PU}
\end{figure}

Jet calibration is a challenging task due to time-dependent changes in both the detector response and calibration and high pileup (PU), which are additional particles originating from secondary proton-proton interactions in the same and neighboring bunch crossings.
During Run 2, on average 29 PU interactions per bunch-crossing were observed (Figure \ref{fig:run2_PU}).
Several techniques, both at event-level and jet-level, can be used to limit the impact of PU on jet energy scale and resolution.
An overview of jet reconstruction procedure, PU mitigation methods, and the jet calibration sequence is presented in the following.

\section{Jet Reconstruction}

\subsection{Event Reconstruction}

\begin{figure}[!ht]
    \centering
    \includegraphics[scale=0.2]{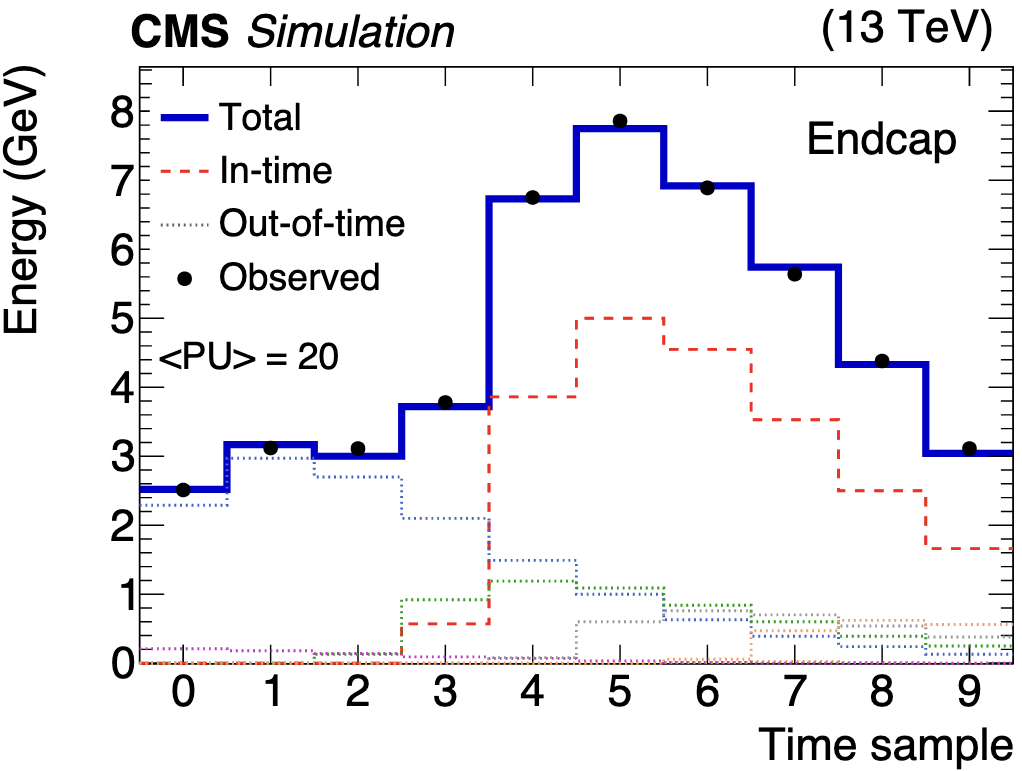}
    \includegraphics[scale=0.277]{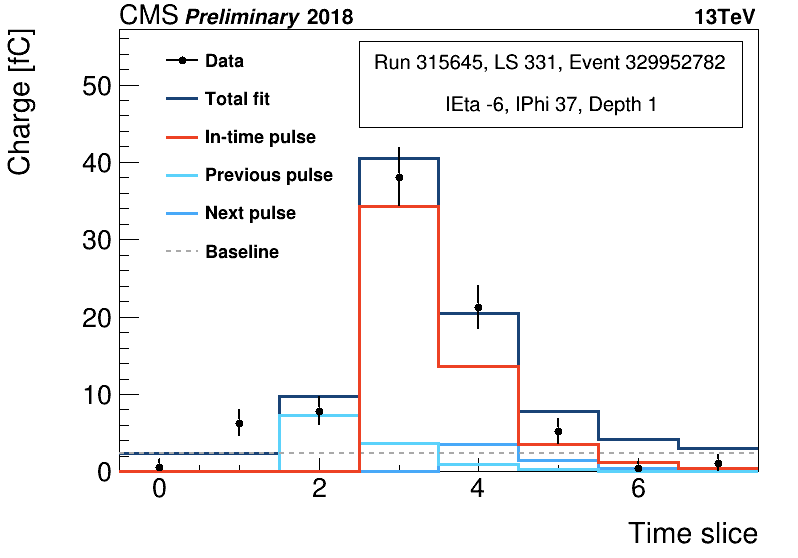}
    \caption{Single channel reconstruction in ECAL \cite{Sirunyan:2721995} and HCAL \cite{CMS-DP-2018-018}. The dots are the digitized data samples, red distribution is the fitted in-time pulse, and light blue distributions are fitted out-of-time pulses.}
    \label{fig:OOTPU}
\end{figure}

Particles produced in proton-proton collisions pass through the CMS detector leaving hits in the tracking system and depositing energies in the electromagnetic and hadronic calorimeters (ECAL and HCAL respectively).
The hits in the tracker are seeded, built using pattern recognition, and fitted to recover the trajectory of the charged particles.
In the calorimeters, the energy deposits are reconstructed as pulses, where the amplitude of the reconstructed pulse corresponds to the measured energy of the particle.
However, due to the finite decay time of the signal in the calorimeters, the total signal contains contributions from the previous and next bunches (Figure \ref{fig:OOTPU}).
Simultaneous pulse shape fitting is performed for both the ECAL and HCAL separately to resolve the signal corresponding to the current in-time pulse and to remove contributions coming from out-of-time pulses. 
The information from the ECAL and HCAL is combined using the Particle Flow (PF) \cite{Dordevic:2678077} algorithm to form clusters. 
A precise calibration is then performed on these calorimeter clusters to correctly reconstruct neutral particles with the right energy scale.
The reconstructed tracks are linked to PF clusters to form charged electromagnetic and hadronic candidates.
PF clusters without linked tracks form neutral hadronic and electromagnetic candidates.
Muons being minimum ionising particles pass through the entire detector and are reconstructed from hits in the inner and outer tracking systems.
At this stage, by combining information from various sub-detectors, a global event description is provided where all final state particles are identified as a charged hadron, neutral hadron, electron, photon or muon candidates.

\subsection{Event-level PU Mitigation}


PU particles produce additional tracks and deposits in the calorimeters which can overlap with that of the jets. 
A majority of PU is from charged particles which can be reduced using the charged hadron subtraction (CHS) method \cite{Dordevic:2678077} , which removes charged particles originating from PU vertices. 
This technique, however, only works within the tracker covered region, and it does not remove neutral PU contribution. 
Another complementary technique is pileup per particle identification, or PUPPI \cite{CMS-PAS-JME-18-001}, where on an event-by-event basis a probability is calculated for each particle describing the degree to which they are pileup-like. These weights are then used
to re-scale the four-momenta of the particles.
As a result, physics objects such as jets and missing energy, and jet substructure variables such as soft-drop mass \cite{Larkoski_2014} and N-subjettiness \cite{Thaler_2011} are expected to be less susceptible to PU when PUPPI is used.

\subsection{Jet Clustering}

At CMS, PF candidates are clustered into jets using the anti-k$_T$ \cite{Cacciari_2008} algorithm which is infrared and collinear safe.
The default PU mitigation methods for Run 2 were to use CHS for narrow jets and PUPPI for large-area jets.
These large-area jets are used in boosted topologies where jet substructure plays an important role. The default for Run 3 is to use PUPPI for both narrow and large area jets.

\section{Jet Calibrations}
CMS follows a factorised approach to calibrating jets which is explained below.
Run 2 legacy reconstruction results shown in Figures \ref{fig:L1Offset} to \ref{fig:JER} are of PF+CHS jets clustered using anti-k$_T$ with R = 0.4.

\subsection{PU Offset Corrections}
\begin{figure}[!ht]
    \centering
    \includegraphics[scale=0.42]{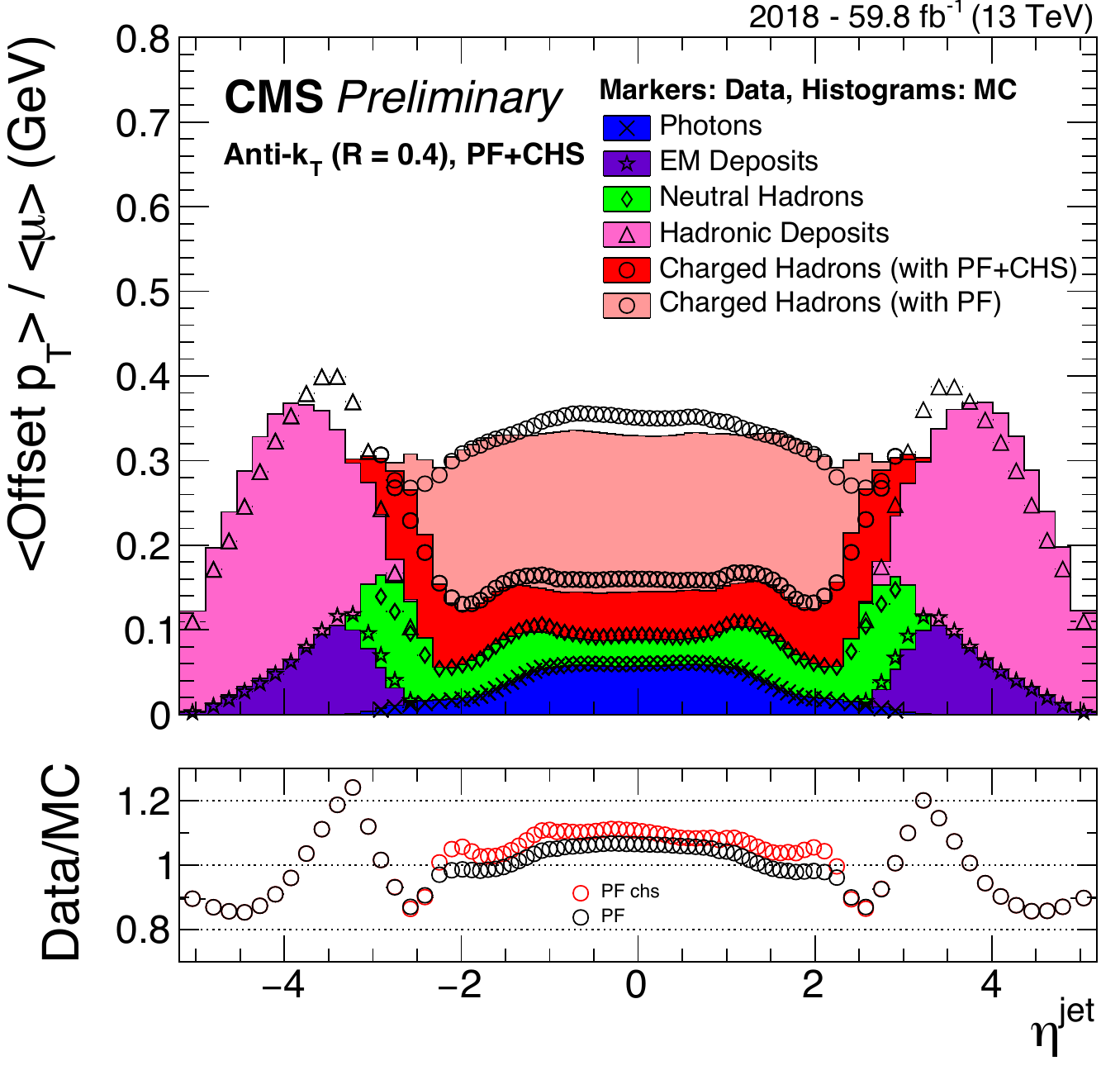}
    \includegraphics[scale=0.42]{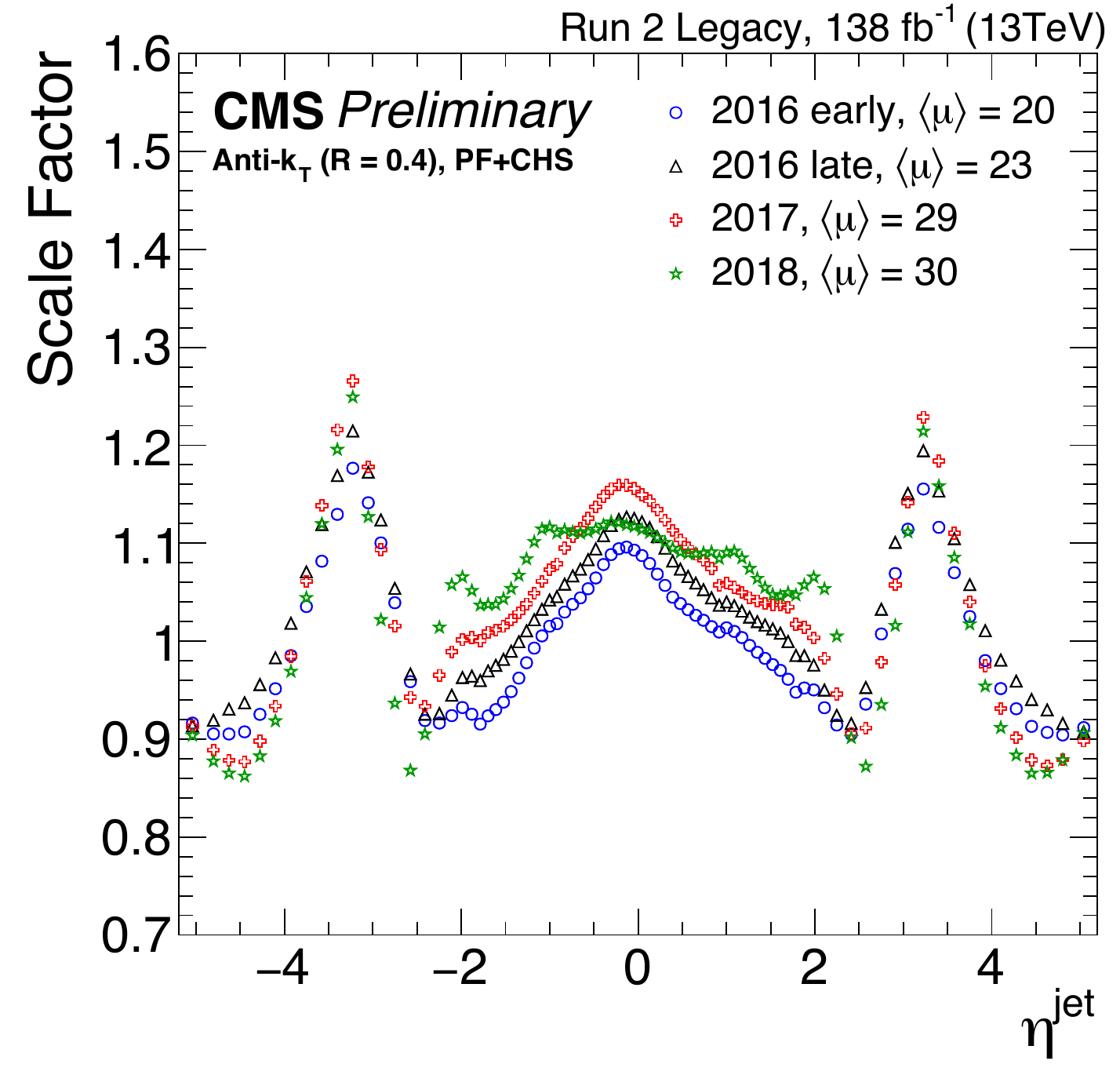}
    \caption{Data-to-simulation comparison for average offset per pileup interaction, calculated for each type of PF candidates (left). Evolution of data-to-simulation scale factors in Run 2 (right). \cite{CMS-DP-2021-033}}
    \label{fig:L1Offset}
\end{figure}

The first step in jet calibration is to estimate and subtract the offset energy coming from PU and noise. 
In simulation, this is done by taking the average difference in transverse momentum (p$_T$) between matched jets, with and without PU overlay, in QCD multi-jet samples and evaluating it as a function of p$_T^{ptcl}$, $|\eta|$, and mean number of pileup interactions per crossing ($\langle\mu\rangle$).
Residual offset corrections for data are derived using the random cone method which takes the average of PF candidate momenta in a randomly placed cone in zero-bias data and simulated samples.
The offset contributions from different PF candidates as a function of $\eta$ for data and simulation are shown in Figure \ref{fig:L1Offset} (left).
The light red fraction corresponds to the charged hadrons associated to pileup vertices that are removed by the CHS algorithm. The data-to-simulation scale-factors for the offset residual corrections are shown in Figure \ref{fig:L1Offset} (right).

\subsection{Simulated Response Corrections}
\begin{figure}[!ht]
    \centering
    \includegraphics[scale=0.42]{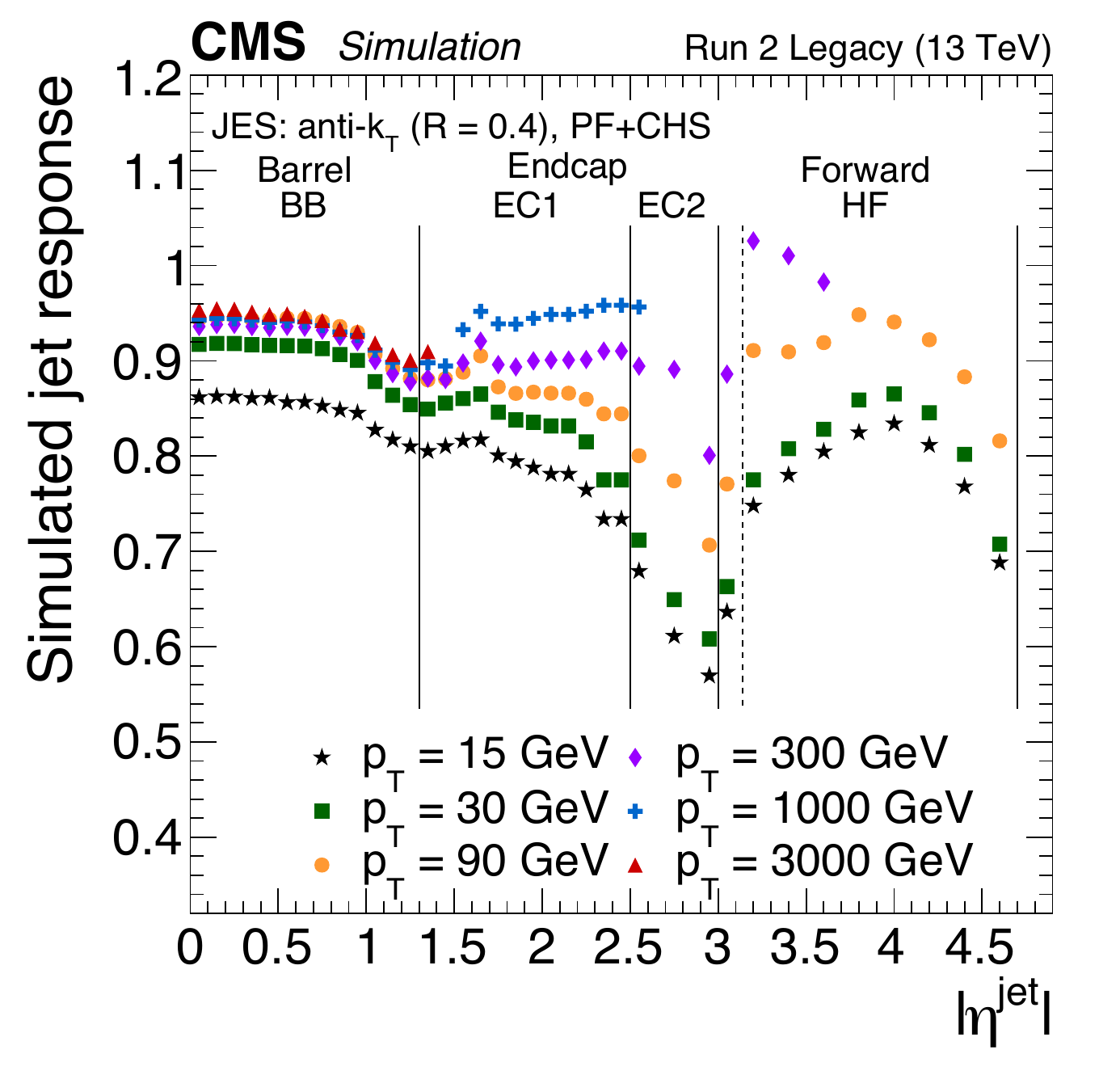}
    \includegraphics[scale=0.42]{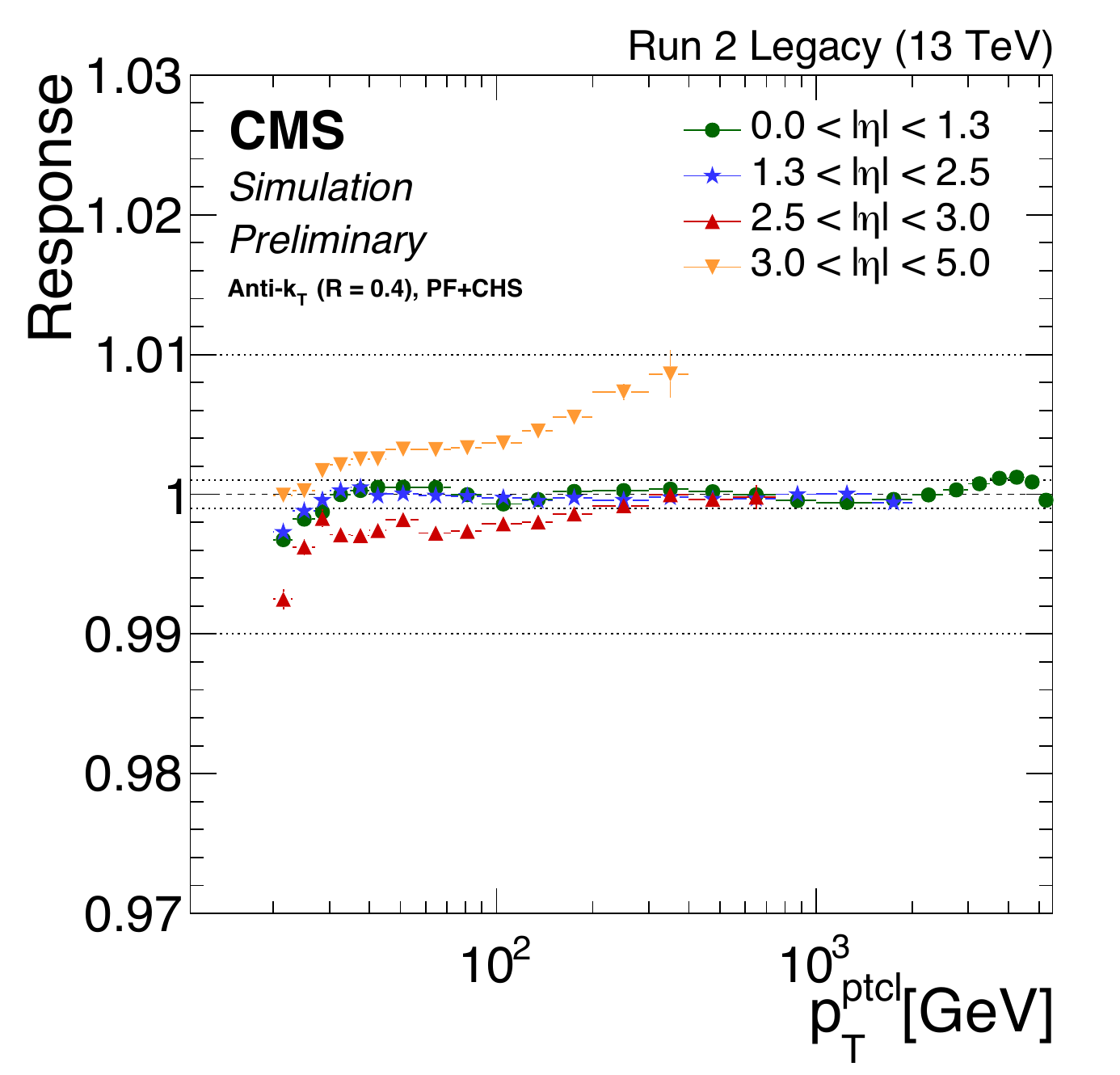}
    \caption{Jet energy response before (left) and after (right) jets are corrected for JES \cite{CMS-DP-2021-033}. From the right plot, we see that the closure is within 0.1\% for $|\eta| <$ 2.5 and within 1\% everywhere else.}
    \label{fig:MCTruth}
\end{figure}

The simulated response corrections account for detector non-uniformity and are derived from PU corrected jets.
The simulated jet response is the ratio of the reconstructed jet p$_T$ to the gen-level p$_T$ (Figure \ref{fig:MCTruth}).
The response is stable in the barrel region and is worse in the detector transition region.
Once this effect is accounted for the closure is within 1\% and 0.1\% for the entire detector as shown in the right panel of Figure \ref{fig:MCTruth}.

\subsection{Residual Corrections for Data}
The residual corrections for data, to account for small differences between simulation and data, are calculated in two steps: relative $\eta$-dependent corrections and absolute p$_T$-dependent corrections.

\begin{figure}[!ht]
    \centering
    \includegraphics[scale=0.42]{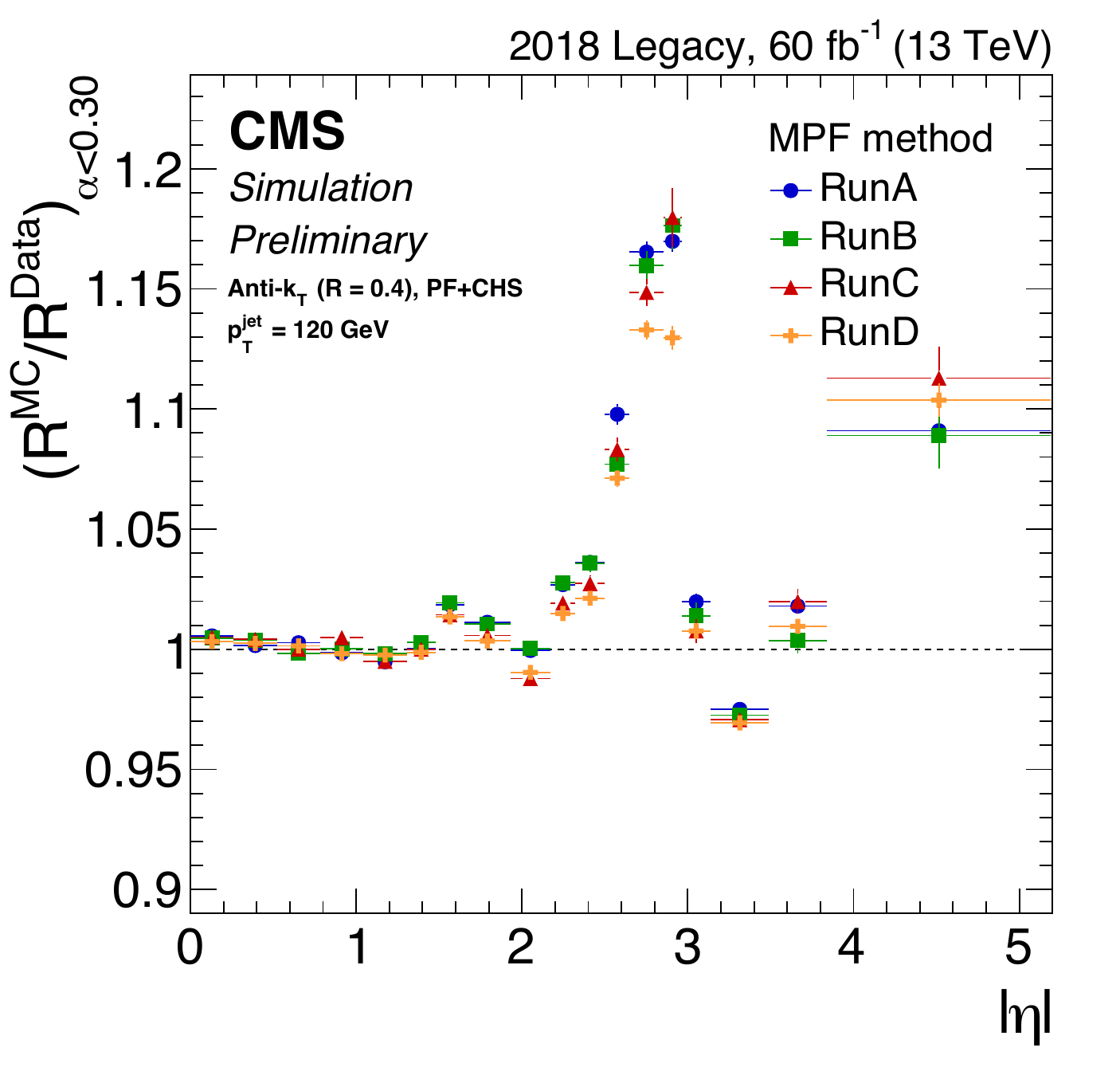}
    \includegraphics[scale=0.42]{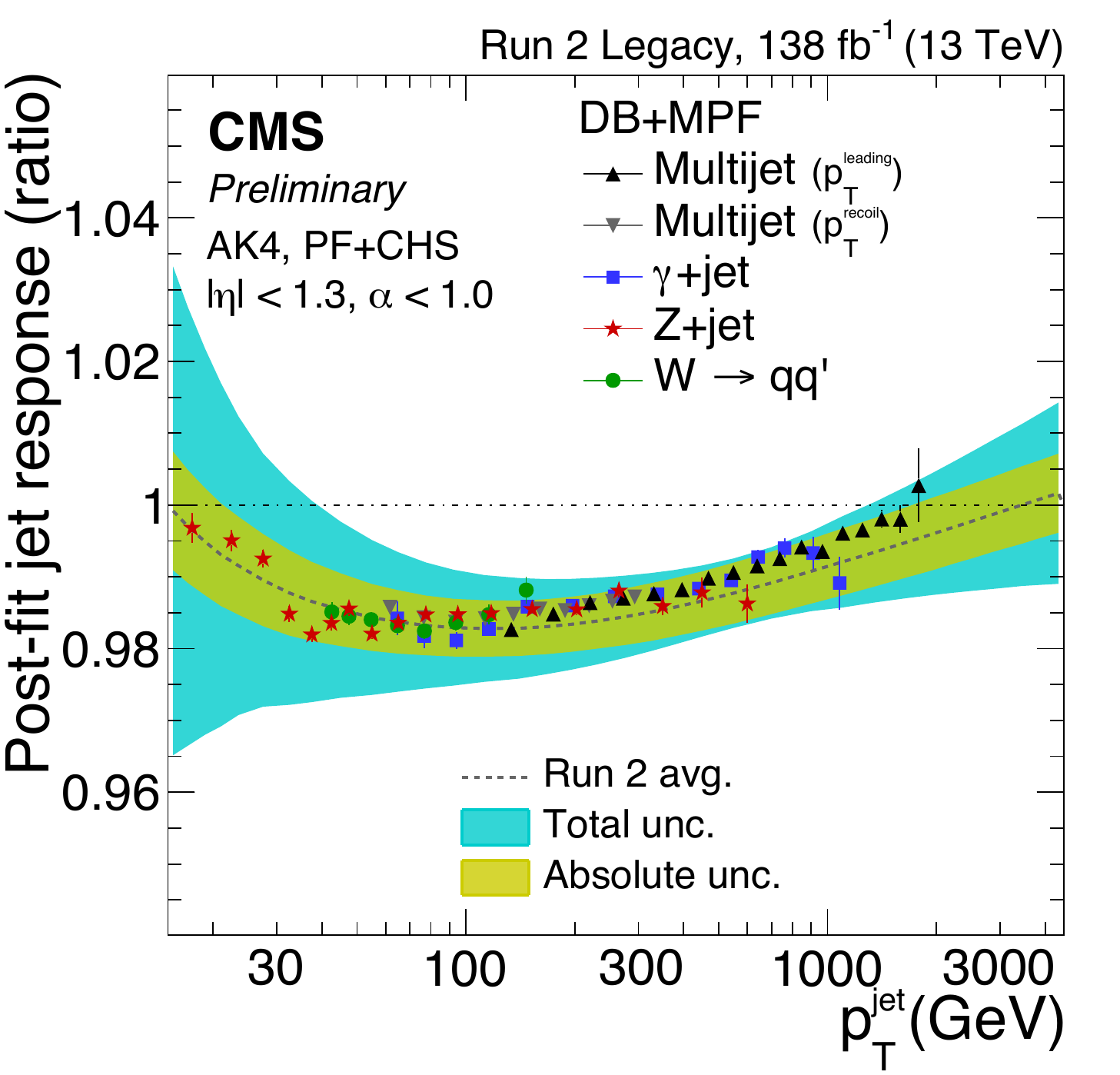}
    \caption{Residual correction derived in bins of $\eta^{jet}$ and p$^{jet}_T$ using dijet events with MPF method (left). Global fit of absolute corrections vs. p$^{jet}_T$ (right). \cite{CMS-DP-2021-033}}
    \label{fig:L2L3Res}
\end{figure}

The \textbf{$\boldsymbol{\eta}$-dependent corrections} are relative corrections where the response of jets in any $\eta$ is corrected with respect to the jet in the barrel region.
As we can see from the left plot in Figure \ref{fig:L2L3Res}, the $\eta$ dependent corrections are less than 5\% everywhere and become sizeable only in the detector transition region.

The \textbf{p$\boldsymbol{_T}$-dependent corrections} exploit the transverse momentum balance between a jet to be calibrated and a precisely calibrated reference object. 
These corrections are derived using different samples and methods covering the entire p$_T$ range.

Several improvements to the global fit (right panel, Figure \ref{fig:L2L3Res}) in Run 2 are:
\begin{enumerate}
    \item The $\alpha$ parameter limits additional jet activity in the event. For instance, in dijet events $\alpha  = p_{T, j_3}/p_{T,avg}$ with similar definitions in other events samples.
    Raising the default value of $\alpha < 0.3$ to $ < 1.0$ removes almost all bias on event PU profile and increases the low p$_T$ statistics in Z+jet samples.
    \item The legacy global fits employ new channels such as using the p$_T$ of the recoil in multijet samples to cover the low p$_T$ regions and using the W+jets samples to cover the p$_T$ region between 40-200 GeV.
    \item The PF composition information is used, bringing the number of fitting parameters to 9 in comparison to the 3 parameter fit used in end-of-year reconstruction. \cite{CMS-DP-2021-033}.
\end{enumerate}

\subsection{Jet Energy Resolution}
\begin{figure}[!ht]
    \centering
    \includegraphics[scale=0.42]{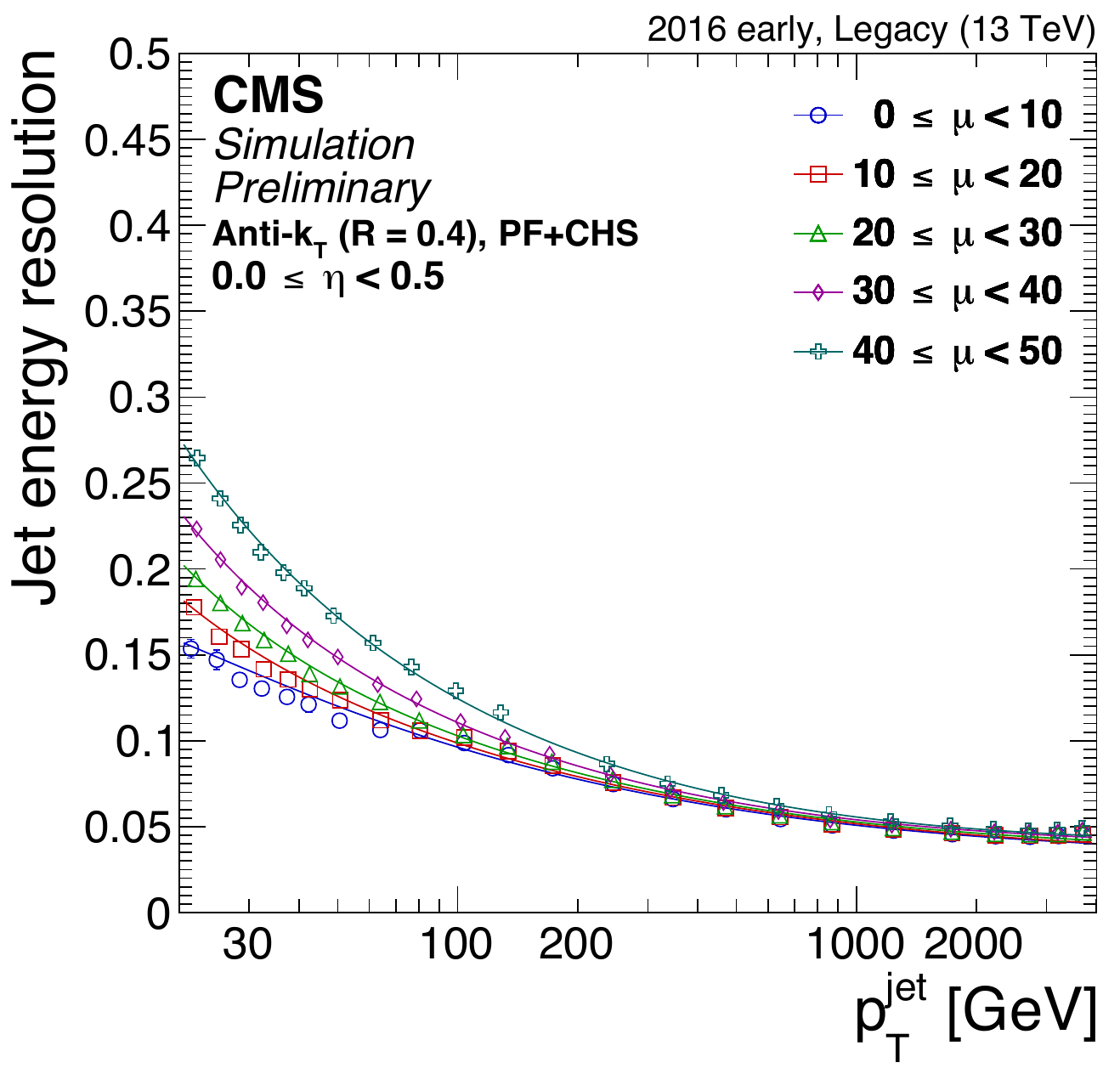}
    \includegraphics[scale=0.43]{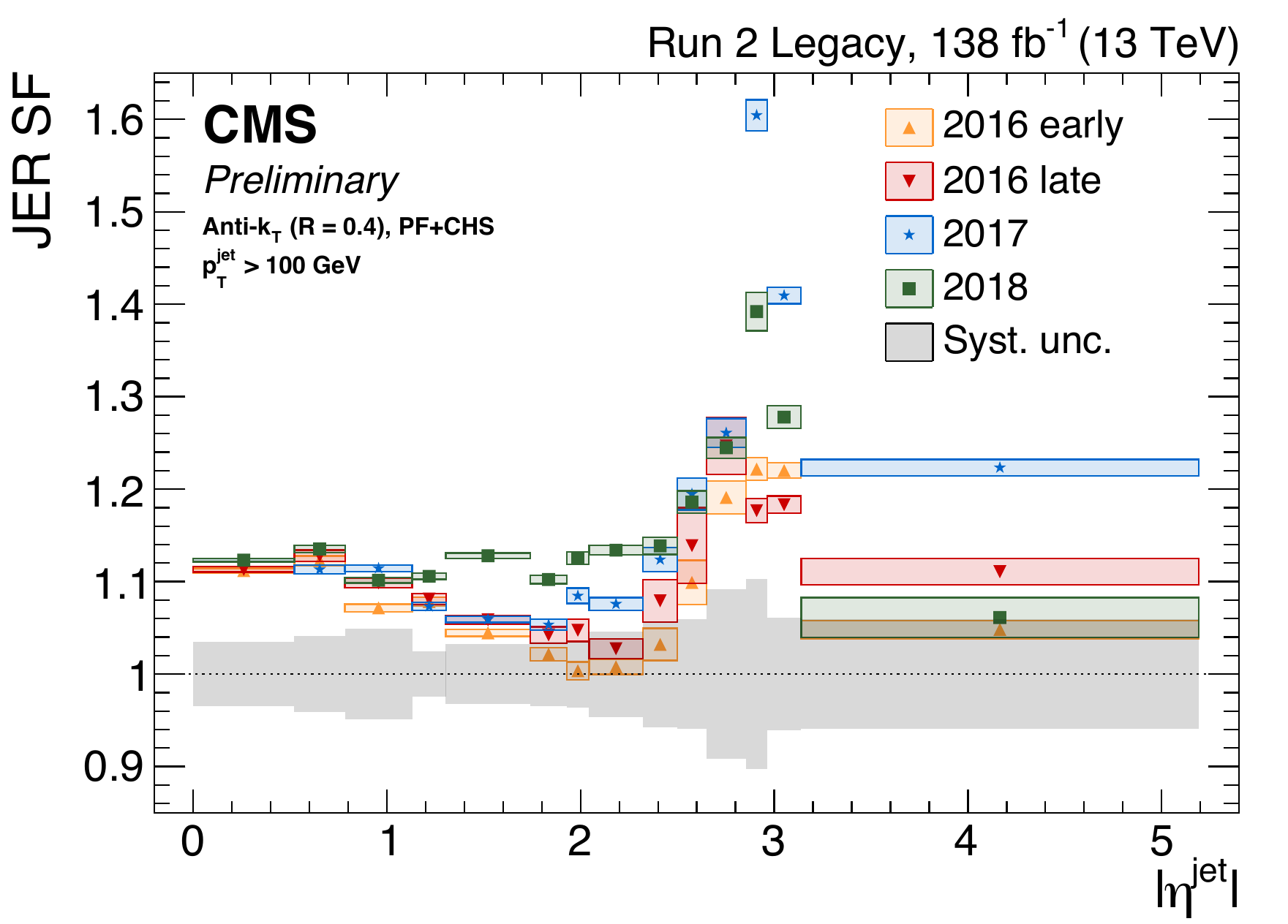}
    \caption{JER versus p$_T$ for varying levels of PU, $\mu$ (left). JER data-to-simulation scale-factors vs. $|\eta^{jet}|$ with the statistical uncertainty shown at each point, while the total systematic uncertainty (gray band) shown around 1. \cite{CMS-DP-2021-033}}
    \label{fig:JER}
\end{figure}

Once the jet energy scale has been corrected, the jet energy resolution in simulation needs to be smeared to match that of data.
The JER in simulation is estimated as the width of the Gaussian fit to the particle-level response for different PU scenarios as a function of jet p$_T$ (left panel, Figure \ref{fig:JER}).
The data-to-simulation scale-factors (right panel, Figure \ref{fig:JER}) are extracted using data based methods and are used to smear the simulated jet resolution.
The scale-factor uncertainties have been reduced significantly for the legacy reconstruction in comparison to the end-of-year reconstruction \cite{CMS-DP-2020-019} and the time-dependence effects can be attributed to residual detector miscalibrations.
For all eras in Run 2, the scale-factors are between 10-15\% and are higher in the detector transition region. 

\section{Summary}
The derivation of JES and JER corrections using various advanced techniques are presented. High PU and evolving detector pose a challenge for the proper calibration of jets. For Run 3, to reduce the impact of PU on jets the PUPPI is used by default.

\nocite{*}
\bibliographystyle{IEEEtran}
\bibliography{refs.bib}

\end{document}